\documentclass[lettersize,journal]{IEEEtran}
\usepackage{amsmath,amsfonts}
\usepackage{algorithmic}
\usepackage{algorithm}
\usepackage{array}
\usepackage[caption=false,font=footnotesize]{subfig}
\usepackage{textcomp}
\usepackage{stfloats}
\usepackage{url}
\usepackage{verbatim}
\usepackage{graphicx}
\usepackage{cite}
\usepackage{booktabs}
\usepackage{makecell}
\usepackage{tabularx}
\usepackage{bm}
\usepackage{xcolor}
\begin{document}
\title{BitSemCom: A Bit-Level Semantic Communication Framework \\ with Learnable Probabilistic Mapping}

\author{Haoshuo Zhang, Yufei Bo, Jianhua Mo and 
        Meixia Tao,~\IEEEmembership{Fellow,~IEEE}
\thanks{
This work is supported by the NSC of China under grant 62125108, by the National Key Laboratory for Positioning, Navigation and Timing Technology, and by the National Sci \& Tech Major Project - Mobile Information Networks under grant 2024ZD1300700.

The authors are with the School of Information Science and Electronic Engineering, Shanghai Jiao Tong University, Shanghai 200240, China (e-mails: \{zhuiguang, boyufei01, mjh, mxtao\}@sjtu.edu.cn). The corresponding author is Meixia Tao. {The source code is available at https://github.com/bhzhs/BitSemCom.}}}



\maketitle

\begin{abstract}
Most existing semantic communication systems based on joint source-channel coding (JSCC) employ analog modulation and are thus inherently incompatible with modern digital communication systems and impose stringent hardware design challenges. Although several digital transmission approaches have been proposed to address this issue, they often suffer from high sensitivity to bit errors, limited adaptability to varying source distributions, or re-training overhead under different modulation schemes. This letter proposes BitSemCom, a novel end-to-end bit-level JSCC framework that is robust to channel noise and modulation-agnostic. The core component is a learnable bit mapper that establishes a probabilistic mapping between continuous {semantic} features and discrete bit {sequences}. By leveraging a sampling-based bit generation method based on the Gumbel–Softmax trick, the framework enables differentiable bit-level optimization while maintaining robustness to channel errors.
Simulation results on image transmission demonstrate that {BitSemCom achieves consistent peak signal-to-noise ratio (PSNR) gains of 2--3 dB over codebook-based digital semantic transmission methods and competitive performance with stronger robustness compared to separate source-channel coding (SSCC) benchmarks.} Ablation studies further validate the effectiveness of the learnable bit mapper.
\end{abstract}


\section{Introduction}
\IEEEPARstart{S}{emantic} communication has emerged as a promising paradigm for next-generation wireless systems. It focuses on extracting and transmitting the essential meaning of source data rather than raw bits. Empowered by advances in artificial intelligence (AI), recent studies have demonstrated the significant advantages of semantic communications on transmission efficiency and task performance, especially in low signal-to-noise ratio (SNR) regimes \cite{Zhangpingzongshu,DeepSC}. A pioneering work in this field is the deep learning-based joint source-channel coding (DeepJSCC) framework  for image transmission \cite{DeepJSCC}. DeepJSCC employs a neural network (NN)-based semantic encoder to map image pixels to continuous-valued channel input signals and a NN-based semantic decoder to reconstruct the image from the noisy received signals.  This framework is categorized as signal-level joint source-channel coding (JSCC). Despite its theoretical appeal, \textit{signal-level} JSCC relies on analog modulation, which is inherently incompatible with the discretized nature of modern digital communication systems, and also imposes stringent challenges on the design of transceiver hardware, such as power amplifiers and automatic gain controllers \cite{Bo-digital}.

\begin{figure}[t]
  \centering
  \includegraphics[width=\linewidth]{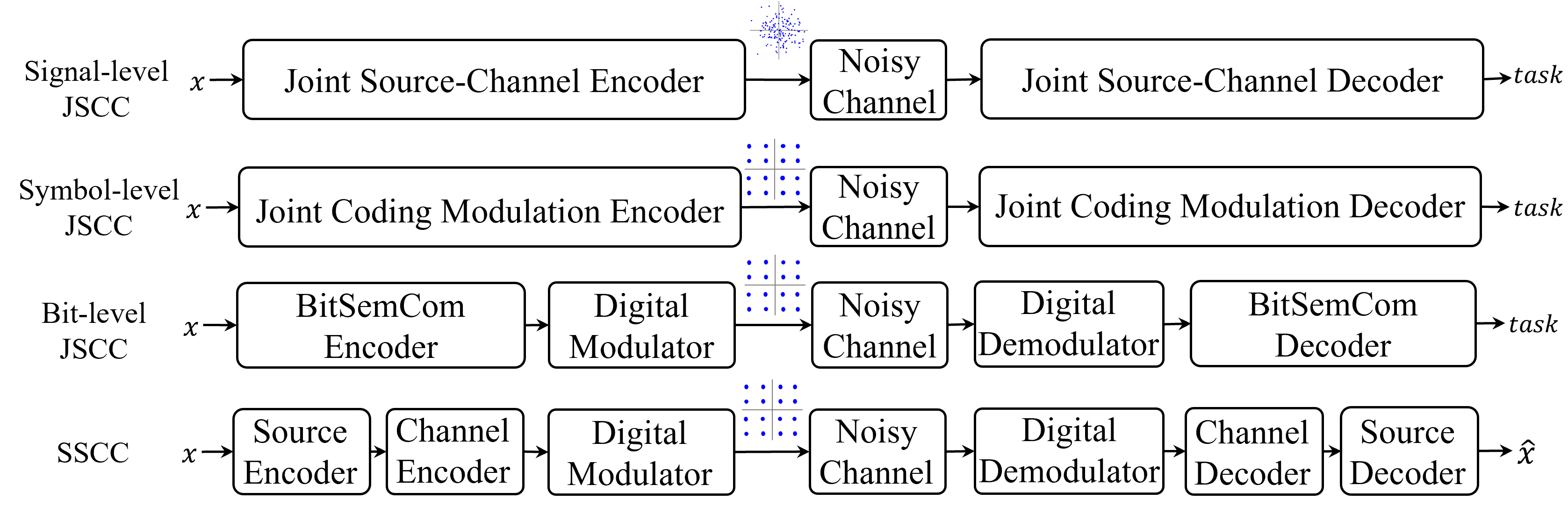}
  \caption{{Framework comparison of different JSCC-based semantic communications (signal-level, symbol-level, and bit-level) and conventional SSCC-based digital communications.}}
  \label{fig:analog_to_digital}
\end{figure}

To address the incompatibility issue and hardware challenges, several digital semantic transmission schemes have been explored, including codebook-based \cite{CodeBook, Bao2025sDAC}, entropy-based \cite{D-JSCC}, and specialized quantizer-based \cite{Quantizer} methods. Specifically, 
the codebook-based schemes employ vector quantized variational autoencoders (VQ-VAEs) \cite{CodeBook} to map semantic features to discrete codewords of a learnable codebook shared by the transmitter and receiver, and then transmit only the corresponding indices. However, decoding is prone to channel-induced errors in such index-based transmission. To enable the learnable codebook to exhibit enhanced robustness against channel noise, the authors in \cite{Bao2025sDAC} propose a semantic digital-to-analog converter (sDAC). This is achieved by employing group-convolutional adapters for feature mapping and incorporating an abstract binary symmetric channel (BSC) into the end-to-end training. 
The entropy-based approaches rely on hyperprior models to estimate the distribution of semantic features and generate bitstreams via entropy coding.
The specialized quantizer-based methods map continuous features into discrete levels via prescribed quantization criteria, which may limit adaptability to varying source distribution and lead to sub-optimal performance.

Different from the digital schemes discussed above, the joint coding-modulation (JCM) framework proposed in \cite{Bo-digital} implements \textit{symbol-level} JSCC. This approach bypasses traditional bit conversion by directly mapping source data onto discrete modulation symbols for transmission using Gumbel-Softmax reparameterization trick \cite{Jang2017GumbelSoftmax}. Such design enables joint optimization of encoding and modulation and supports end-to-end training over wireless channels.
While this symbol-level JSCC scheme offers better
compatibility than signal-level JSCC {by explicitly generating digital modulated symbols rather than continuous-valued signals,} it still faces one major
limitation. That is, the mapping is designed for a specific
modulation scheme (such as rectangular 16QAM) and, as a
result, the model needs retraining whenever the modulation
order or constellation shape changes.



In this letter, we propose \textbf{BitSemCom}, an end-to-end \textit{bit-level} JSCC framework that can seamlessly reuse existing modulation, waveform, and beamforming techniques and hence is easily compatible with current digital devices and air-interface protocols.
{An illustrative framework-level comparison of bit-level JSCC with prior JSCC and the conventional separate source-channel coding (SSCC) is shown in  Fig.~\ref{fig:analog_to_digital}.}
The core of the framework is a learnable probabilistic bit mapper, which establishes a differentiable bridge between continuous semantic features and discrete bit sequences.  By incorporating the Gumbel-Softmax trick, our framework enables the system to optimize bit-generation probabilities directly through backpropagation. 
Therefore, the model is able to learn an error-resilient bit representation that is adapted to both source distributions and channel statistics. Unlike symbol-level JSCC, our bit-level JSCC approach is modulation-agnostic, maintaining high performance across various digital modulation formats without requiring architecture modifications or re-training.

Extensive simulation results on image transmission demonstrate that {BitSemCom achieves consistent peak signal-to-noise ratio (PSNR) gains of 2–3 dB over codebook-based digital semantic transmission methods. It also shows stronger robustness than conventional SSCC benchmarks.} Ablation studies further validate the effectiveness of the learnable bit mapper.

\begin{figure}[t]
  \centering
  \includegraphics[width=\linewidth]{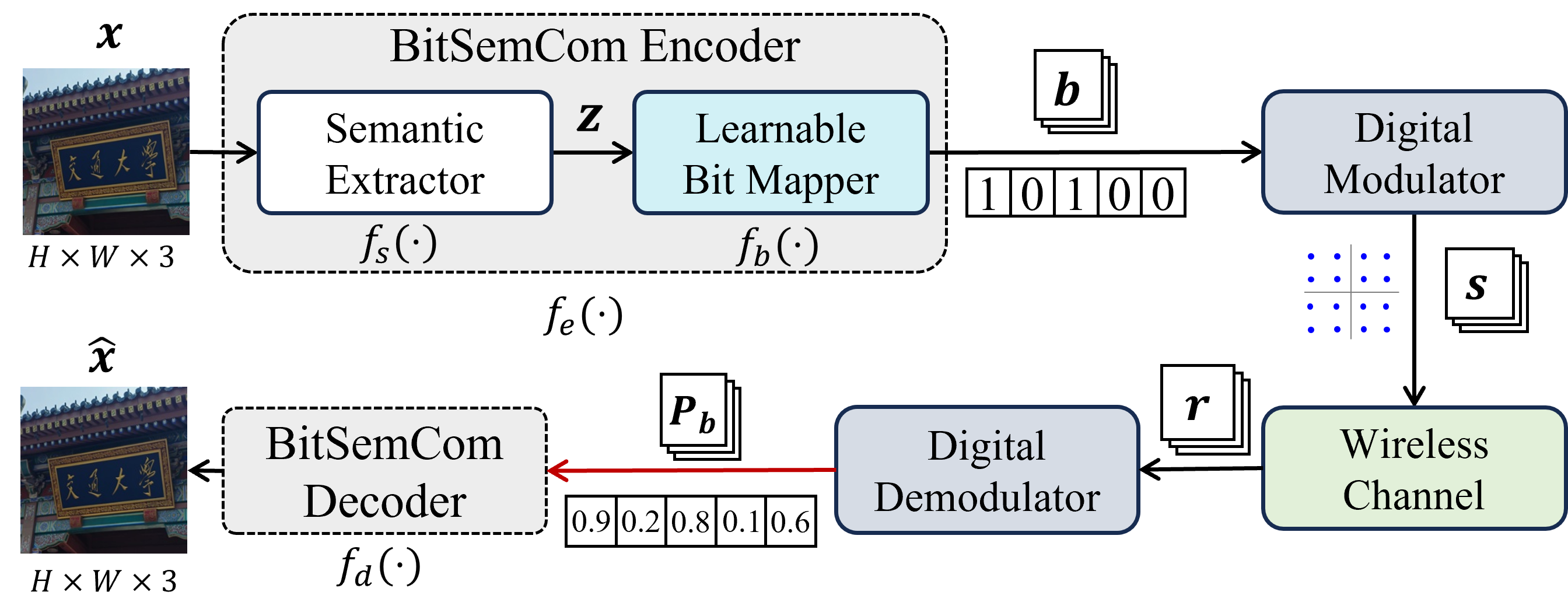}
  \caption{{Overall framework of the BitSemCom.}}
  \label{fig:system}
\end{figure}

\section{Overall Framework of BitSemCom}
We consider an end-to-end image transmission system, as illustrated in Fig. \ref{fig:system}. The input image is denoted as $\bm{x} \in \mathbb{R}^{H \times W \times 3}$, where $H$ and $W$ represent the height and width of the image, respectively. 
{At the transmitter, the input image is encoded by a BitSemCom encoder consisting of a semantic extractor $f_s(\cdot)$ and a learnable bit mapper $f_b(\cdot)$. The semantic extractor first produces a continuous semantic representation $\bm{z} \in \mathbb{R}^{L_z}$ of dimension $L_z$. The bit mapper then converts $\bm{z}$ into a binary sequence $\bm{b} \in \{0,1\}^{L_b}$ with length $L_b$.}\footnote{Here, $L_z$ and $L_b$ can be different. For a fair comparison with baseline methods, we set $L_z = L_b$ in this letter.} The detailed bit generation mechanism will be elaborated in Section~III.
The bit sequence $\bm{b}$ is then mapped into a sequence of complex-valued symbols $\bm{s} \in \mathbb{C}^{L_s}$ via any given digital modulation (e.g., $M$-QAM), where the symbol length is determined by $L_s=\frac{L_b}{\log_2M}$. These symbols are normalized to satisfy the unit average power constraint, i.e., $\frac{1}{L_s}\mathbb{E}[||\bm{s}||^2] = 1$. Finally, the modulated symbols are transmitted using $L_s$ channel uses, with the transmission efficiency evaluated by channel uses per pixel (cpp), defined as $k = L_s / (H \times W)$. 

At the receiver, the received signal $r_i$ at each channel use $i$ is given by:
\begin{equation}
    r_i = h\sqrt{p}s_i + n_i, \quad i=1, 2, \ldots, L_s,
\end{equation}
where $h \in \mathbb{C}$ denotes the channel coefficient, $p$ represents the transmission power, and $n_i \sim \mathcal{CN}(0, \sigma_n^2)$ is the additive white Gaussian noise (AWGN). {Upon receiving $\bm{r}=[r_1,r_2,\ldots,r_{L_s}]$, the demodulator performs soft-decision to obtain the log-likelihood ratios (LLRs) for each element in the bit sequence $\bm{b}$, which are then converted into a bit-wise posterior probability vector $\bm{P_b} = [P_1, P_2, \ldots, P_{L_b}]$. Here, $P_i = P(b_i = 1 \mid \bm{r})$. Finally, a BitSemCom decoder $f_d(\cdot)$ reconstructs the image $\bm{\hat{x}} \in \mathbb{R}^{H \times W \times 3}$ from $\bm{P_b}$.}




\section{Learnable Bit Mapper}

In this section, we present the key component of the proposed BitSemCom, namely a differentiable learnable bit mapper that transforms continuous semantic features into discrete bitstreams in a probabilistic manner. 
{Note that an early version of the bit mapper is adopted in our previous work for multi-spectral image segmentation \cite{ProMSC}. While it serves merely as a digitization tool for source coding in \cite{ProMSC}, the bit mapper in this work also enables error protection by explicitly accounting for channel impairments in a bit-level JSCC framework. In addition, whereas the bit mapper was mentioned briefly at the functional level in \cite{ProMSC}, this letter offers a much deeper and more thorough investigation from the perspective of architecture design.}

\begin{figure}[t]
  \centering
  \includegraphics[width=\linewidth]{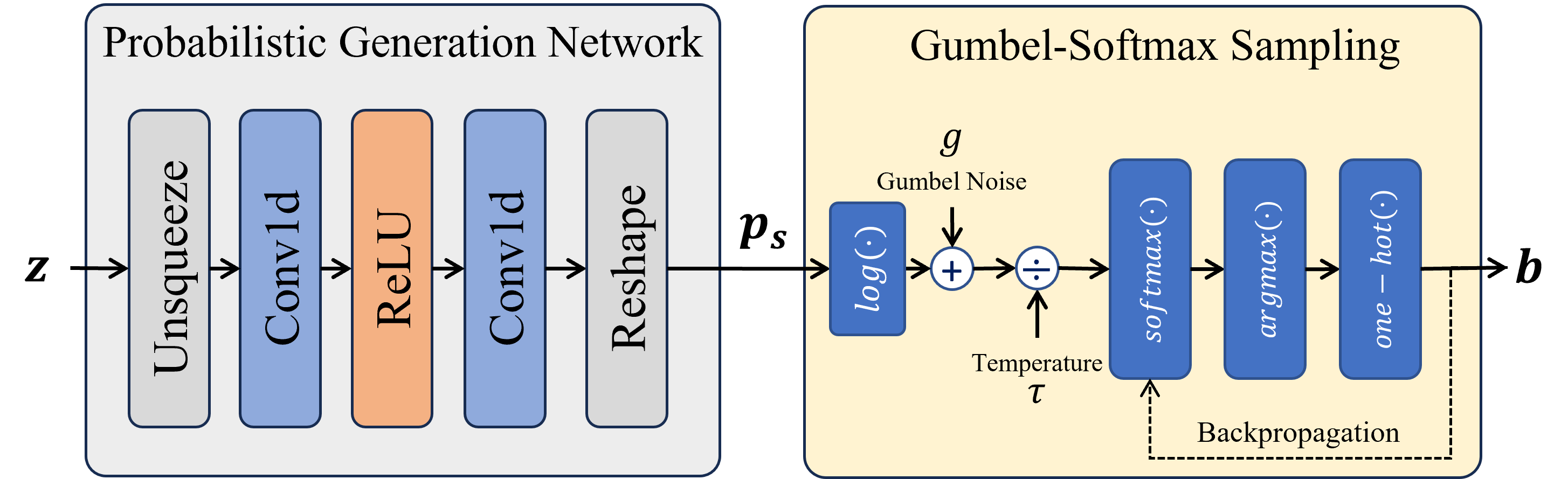}
  \caption{{The structure of the learnable bit mapper.}}
  \label{fig:Bitmapper}
\end{figure}

As illustrated in Fig.~\ref{fig:Bitmapper}, the proposed bit mapper consists of two modules: a probabilistic generation network and a Gumbel-Softmax sampling layer.
The probabilistic generation network employs one-dimensional (1D) convolutional layers to capture local dependencies along the feature dimension. Conditioned on the continuous feature vector $\bm{z}$, it produces a probability distribution $\bm{p_s}=[\bm{\pi_1},\bm{\pi_2},\ldots,\bm{\pi_{L_b}}]^\top \in \mathbb{R}^{L_b\times2}$, where the $\ell$-th row $\bm{\pi}_\ell=[\pi_\ell^0,\pi_\ell^1]$ represents the probabilities of bit $b_\ell$ being 0 and 1. This formulation characterizes the bitwise probability distributions over all $L_b$ bits.

To enable discrete bit generation during forward propagation, we adopt the Gumbel-Softmax reparameterization trick as in \cite{Bo-digital}. This technique allows us to generate discrete bit sequences during forward propagation and estimate the gradients of the probabilistic generative network during backpropagation.

During forward propagation, the Gumbel-Max sampling method reparameterizes the sampling process by introducing Gumbel noise.
For each bit $b_i$, a one-hot vector $\bm{t_i}\in \{0,1\}^2$ is sampled from $\bm{\pi_i}$ as 
\begin{equation}
\bm{t_i} = \operatorname{one\text{-}hot}\left(
\arg\max_{m \in \{0,1\}}
\big( g_i^m + \log(\pi_i^m) \big)
\right),
\end{equation}
where $g_i^m \sim \text{Gumbel}(0,1)$ are independent and identically distributed (i.i.d.) samples from the standard Gumbel distribution. The discrete bit is then obtained as $b_i = \mathbf{e}^\top \bm{t_i}$ with $\mathbf{e} = [0,1]^\top$, converting the one-hot vector into a binary value. The resulting bit sequence $\bm{b}$ is generated according to the learned probability distribution $\bm{p_s}$.

During backpropagation, \text{argmax} function of Gumbel-Max method is approximated by the differentiable \text{softmax} function, thus $t_i$ is substituted by a vector $\bm{\tilde{t}_i}=[\tilde{t}_{i}^0,\tilde{t}_{i}^1]$, i.e.,
\begin{equation}
   \tilde{t}_i^{m} = \frac{\exp((\log(\pi_i^m)+g_i^m)/\tau)}{\sum_{l=0}^1\exp((\log(\pi_i^l)+g_i^l)/\tau)}, m =0,1,
\end{equation}
where $\tau>0$ is a temperature parameter that controls the smoothness of the approximation. As $\tau$ decreases, $\bm{\tilde{t}_i}$ approaches the one-hot vector $\bm{t_i}$ while remaining differentiable for gradient-based optimization.

Before concluding this section, we would like to emphasize that the proposed learnable bit mapper which maps the semantic feature $\bm{z}$ of dimension $L_z$ to the  bit sequence $\bm{b}$ of the length $L_b$ is not a conventional quantizer, but a differentiable binary sampling module. By taking channel conditions into account during the optimization of the bit mapper, the module acquires resilience to channel noise. {Moreover, the BitSemCom decoder directly reconstructs the source image from posterior probability sequences, without requiring de-quantization or latent feature recovery at the receiver.} Together, this unified design enables BitSemCom to operate as a fully integrated end-to-end bit-level JSCC framework.

\section{Experiment Results}
\subsection{Training Strategies}

The overall training consists of two stages. First, we adopt the pretrained semantic extractor and decoder with Swin Transformer architecture from \cite{WITT}, and train the bit mapper jointly with them over an error-free channel. Second, the system is fine-tuned over a noisy channel to enhance robustness. 
{In the second training stage, we consider both an AWGN channel with a fixed coefficient $h=1$, and a quasi-static block fading channel where $h$ remains constant within each transmission block but varies independently across different transmissions. Here, Rayleigh fading is assumed, and each specific realization of $h$ is available at the receiver so that coherent equalization is performed accordingly. For each training batch, the average SNR is randomly sampled from $-3$ to $6$ dB. Unless otherwise specified, all SNR values refer to the average SNR for both training and test.}


\subsection{Experimental Settings and Benchmarks}

We use the DIV2K dataset \cite{DIV2K} for training and evaluate the task performance on the Kodak dataset \cite{Kodak1993}. During training, images are randomly cropped into $256 \times 256$ patches. The model is trained using Adam with a batch size of 32 and a step-wise learning rate schedule, where the initial learning rate is set to $1\times10^{-4}$ and decayed by a factor of 0.9 every 200 epochs. The mean squared error (MSE) loss is employed for optimization. 

{The semantic extractor and BitSemCom decoder are both built upon a hierarchical Swin Transformer backbone \cite{WITT}. The extractor consists of four stages of Swin Transformer blocks, where the feature dimension is progressively increased while the spatial resolution is gradually reduced via patch merging operations. A linear projection head is then employed to generate the latent representation $\bm{z}\in \mathbb{R}^{\frac{H}{16}\times \frac{W}{16} \times C}$ where $C$ denotes the output embedding dimension of the projection head. The BitSemCom decoder adopts a symmetric architecture. In our experiments, we set $C = 32$ and $C = 64$, corresponding to $L_b = 8192$ and $16384$, respectively. Both QPSK and 16QAM modulation schemes are adopted. Under these settings, two CPP values are obtained, i.e., $k = 1/16$ and $k = 1/8$.}


We consider the following digital schemes as benchmarks:
\begin{itemize}
    \item \textit{BPG/JPEG2000 + LDPC:} This is a traditional non-learning-based SSCC scheme, where the image is first compressed using BPG or JPEG2000 and then protected by low-density parity-check (LDPC) codes for channel transmission.
    \item \textit{JPEG AI + LDPC:} This is a learning-based SSCC method, where the state-of-the-art entropy-based image compression framework JPEG AI \cite{JPEG_AI} is adopted for source coding.
    \item \textit{VQVAE + LDPC:} This is a VQ-VAE codebook-based method \cite{CodeBook}, where the codebook is trained by assuming error-free transmission but the codeword indices are protected by LDPC codes for actual channel transmission.
    \item \textit{sDAC:} This is the codebook-based scheme proposed in \cite{Bao2025sDAC}, where the codebook is learned in an end-to-end manner over a noisy channel.
\end{itemize}

LDPC codes follow the Digital Video Broadcasting – Satellite – Second Generation (DVB-S2) standard \cite{DVB_S2}, with a block length of 16,200 bits and code rates of 1/4, 1/2, and 3/4.  For reference, an ideal case with capacity-achieving channel codes is also included, denoted as “BPG + Capacity”.

Performance is evaluated using PSNR and learned perceptual image patch similarity (LPIPS){\cite{LPIPS}}. PSNR measures pixel-wise reconstruction quality, while LPIPS captures perceptual similarity via deep feature embeddings.

\subsection{Performance Comparison}
\begin{figure*}[t]
  \centering
  \includegraphics[width=\linewidth]{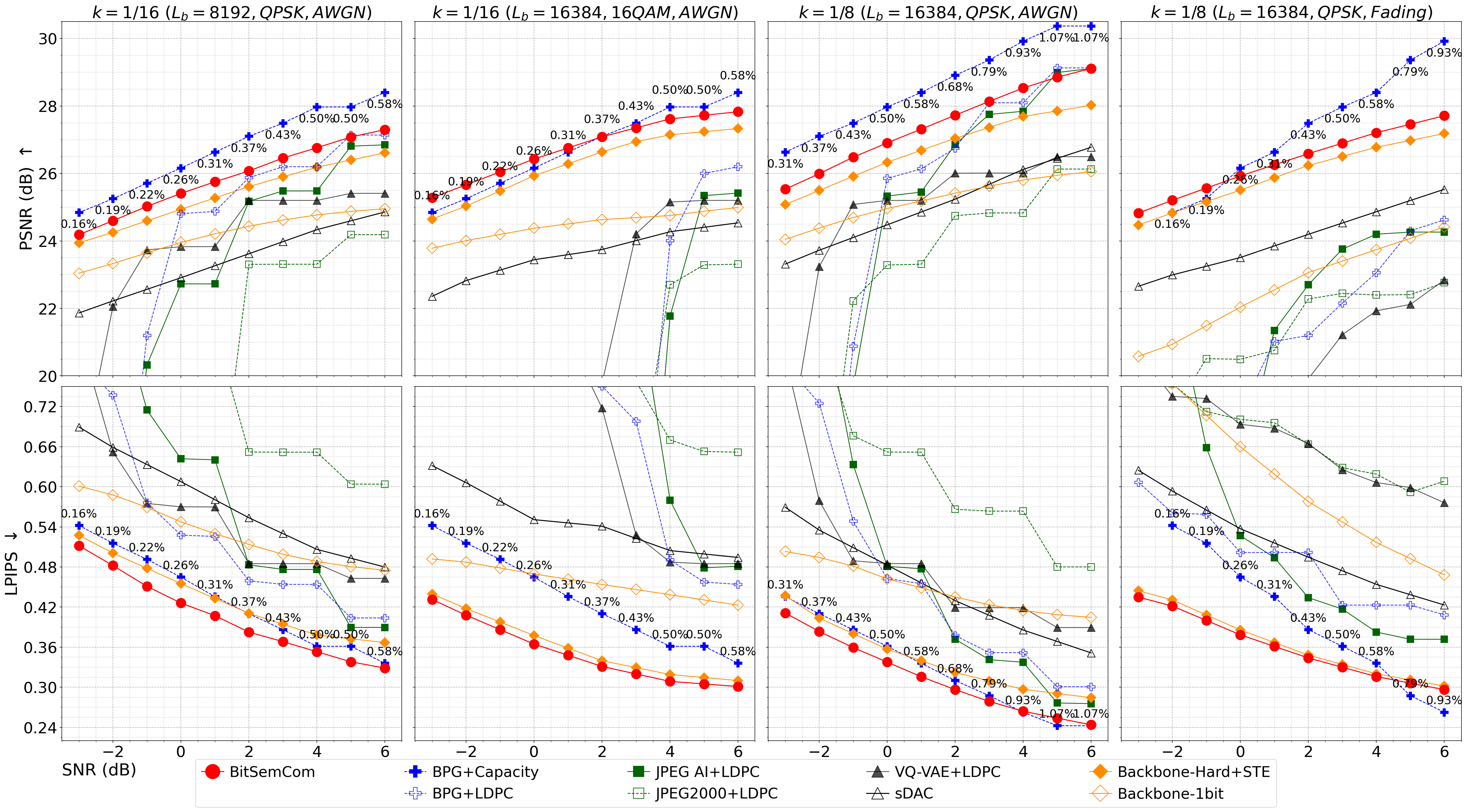}
  \caption{{Performance comparison of BitSemCom against benchmarks and ablation variants. For all the SSCC benchmarks employing LDPC codes or ideal capacity-achieving channel codes, performance envelopes are plotted across different source compression rates and SNR conditions. For BPG + Capacity, different source compression rates are indicated under varying SNR conditions.}}
  \label{fig:performance}
\end{figure*}

Fig.~\ref{fig:performance} compares the proposed BitSemCom with the benchmarks {at different channel uses, modulation schemes, and channel conditions.}
We first observe that, compared to existing codebook-based digital semantic transmission schemes (i.e., sDAC and VQ-VAE+LDPC), BitSemCom demonstrates consistent gains in both PSNR and LPIPS across all tested scenarios. {Specifically, BitSemCom obtains 2--3dB gain in PSNR over sDAC, while the gain over VQ-VAE+LDPC becomes more pronounced at higher modulation and fading channels.} {This performance gain arises from their different representation mechanisms. Codebook-based methods quantize semantic features into a limited vector space using a finite set of discrete codewords, whereas BitSemCom directly maps semantic features to bitstreams, thereby avoiding the performance limitation imposed by the codebook size. These results further validate that BitSemCom captures joint source–channel statistics more effectively than learned codebooks.}
Notably, sDAC only outperforms VQ-VAE+LDPC at low SNRs {under AWGN channels (especially in the QPSK setting),} which can be attributed to the differences in codebook size and training strategy. In our simulations, VQVAE+
LDPC trains its codebook of size 4096 without considering channel
errors. In contrast, sDAC accounts for channel noise during training, and its performance
saturates when the codebook size exceeds 16 \cite{Bao2025sDAC}. Following the original setup, we set sDAC’s codebook size to 16, which may limit its representation capacity and leads to inferior performance at moderate-to-high SNRs.

It is also observed that compared with traditional SSCC schemes (e.g., BPG+LDPC and JPEG AI+LDPC), BitSemCom demonstrates competitive performance with strong robustness to channel noise. 
{For example, at an SNR of 6 dB, BitSemCom achieves PSNR gains of 1.6 dB under 16QAM over AWGN channels and approximately 3 dB under QPSK over fading channels compared with BPG+LDPC. This can be attributed to the fact that SSCC-based methods are highly sensitive to channel variations and often suffer from the rate-mismatch issue. 
In contrast, BitSemCom jointly optimizes source and channel characteristics in an end-to-end manner, and maintains more stable performance across different channel conditions.}

{We further analyze the impact of cpp, modulation order, and channel fading on the performance of BitSemCom. First, increasing the cpp from $k=1/16$ to $k=1/8$ yields a consistent gain of approximately 1.5--1.8 dB in PSNR under QPSK and AWGN settings, demonstrating that the proposed framework can effectively scale with available bandwidth. Second, under the same channel use setting ($k=1/16$), 16QAM provides a moderate gain of around 1.0 dB at low SNRs compared to QPSK due to higher spectral efficiency, while the advantage diminishes at higher SNRs, reflecting the trade-off between modulation efficiency and noise sensitivity. Finally, with fading channels, BitSemCom still preserves a stable trend and consistently outperforms the benchmarks.}


\subsection{Ablation Study}
To evaluate the gain introduced by the proposed learnable bit mapper, ablation studies are conducted:
\begin{itemize}
\item  \textit{Backbone-1bit:} This variant replaces the learnable bit mapper with 1-bit uniform quantization, using uniform noise during training to approximate quantization and a round operation at inference to generate bitstreams.
\item  \textit{Backbone-Hard + STE:} This variant replaces the learnable bit mapper with a sigmoid-based hard-decision module. To address the non-differentiability of hard decision, the STE is employed during training.
\end{itemize}

\begin{figure}[t]
  \centering
  \includegraphics[width=0.9\linewidth]{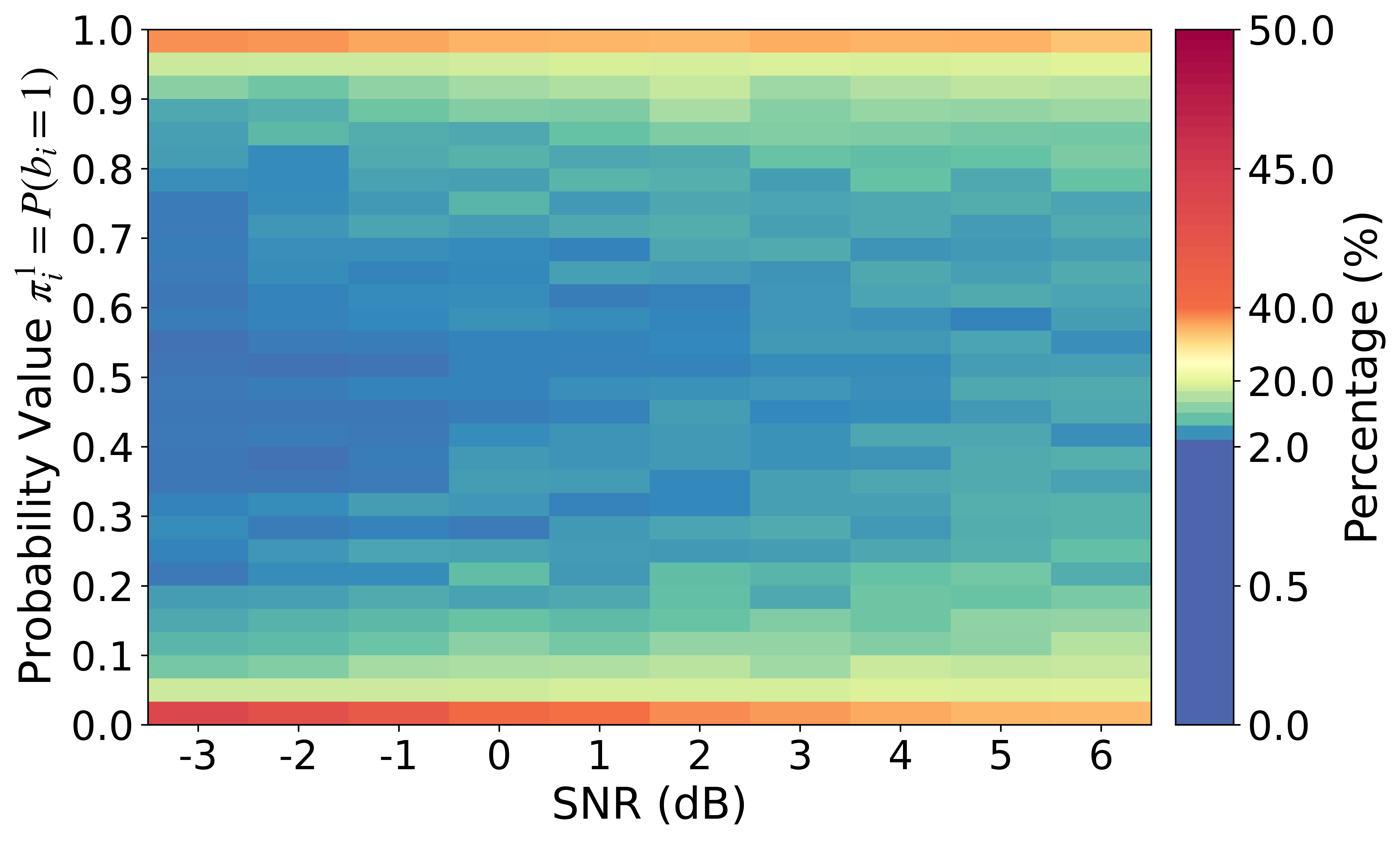}
\caption{{Heatmap of the relative frequency histogram of probability values $\pi_i^1 = P(b_i=1)$ for $i = 1, 2, \dots, L_b$ under different SNRs ($L_b = 16384$, QPSK over AWGN channels).}}
  \label{fig:PDF}
\end{figure}

As illustrated in Fig.~\ref{fig:performance}, BitSemCom consistently outperforms these ablation variants across all tested SNRs, confirming the effectiveness of the learnable bit mapper. Notably, the performance gap between BitSemCom and Backbone-Hard+STE gradually narrows as the SNR decreases, {especially under QPSK modulation and AWGN channels.} To investigate the mechanism behind this trend, we {plot the heatmap of the relative frequency distribution} of bit-sampling probability values $\pi_i^1$ (the probability of $b_i$ being 1) generated by BitSemCom {against} varying SNR conditions in Fig.~\ref{fig:PDF}. {We partition the probability interval $[0, 1]$ into 30 uniform bins, and the color intensity indicates the percentage of bits falling into each bin, using a non-linear color mapping.}
The results reveal a clear adaptive behavior of the proposed bit mapper. At high SNRs, the relative frequency distribution is relatively flat across probability bins, indicating increased randomness in bit generation. This randomness enhances representational flexibility and leads to superior performance compared with Backbone-Hard+STE, which employs a fixed hard-decision mapping and always produces deterministic bits. As the SNR decreases, the relative frequency increasingly concentrates around the probability bins 0 and 1, yielding progressively more deterministic outputs. In low-SNR regimes, BitSemCom therefore converges toward the deterministic behavior of Backbone-Hard+STE, leading to a reduced performance gain. Overall, these observations confirm that the learnable bit mapper adaptively balances stochasticity and determinism according to channel conditions.


\subsection{Storage Overhead and Computational Complexity}

Table~\ref{tab:flops} compares the parameter counts and floating-point operations (FLOPs) across different schemes. 
Both Backbone-1bit and Backbone-Hard+STE adopt non-parametric digitization modules and thus introduce no additional storage or computational overhead. In contrast, sDAC and VQ-VAE introduce additional overhead from their learnable codebooks, with sDAC using a small
codebook of size 16 as in the original paper, while VQ-VAE adopts a much larger one (size 4096). This discrepancy also aligns with their performance gap in Fig.~\ref{fig:performance}. Notably, JPEG AI is excluded from this table because its different backbone architecture would make a direct complexity comparison unfair.

It can be observed that the learnable bit mapper in BitSemCom adds only 0.12M parameters and 0.03G FLOPs, corresponding to negligible increases of 0.43\% and 0.09\%, respectively,  relative to the Backbone.
In contrast, VQ-VAE introduces a much larger overhead, accounting for 6.93\% and 2.92\% increases. 
Despite a slightly higher complexity compared with sDAC, BitSemCom achieves superior task performance as shown in Fig.~\ref{fig:performance}.


\begin{table}[t]
\small
\centering
\caption{Comparison of Parameters and FLOPs}
\label{tab:flops}

\setlength{\tabcolsep}{4pt}
\renewcommand{\arraystretch}{1.25} 
\newcolumntype{C}[1]{>{\centering\arraybackslash}m{#1}} 
\begin{tabularx}{\linewidth}{C{0.42\linewidth}|C{0.23\linewidth}|C{0.23\linewidth}}
\Xhline{1pt}
Model & Params  & FLOPs  \\
\Xhline{1pt}
Backbone / Backbone-1bit, Backbone-Hard+STE & 28.13M& 33.59G \\
\hline
sDAC & +232 & +1.11M \\
\hline
VQ-VAE & +1.95M & +0.98G \\
\hline
BitSemCom & +0.12M & +0.03G \\
\Xhline{1pt}
\end{tabularx}
\end{table}

\section{Conclusion}

In this letter, we propose BitSemCom, a fully integrated end-to-end bit-level JSCC framework that employs a probabilistic sampling-based bit generation mechanism for semantic communications. Experimental results show that BitSemCom consistently outperforms codebook-based digital semantic transmission methods in PSNR by 2–3dB over a wide range of system parameters. It also exhibits competitive performance with stronger robustness compared to conventional SSCC benchmarks.

\bibliographystyle{IEEEtran}
\bibliography{BitSemCom}

\vfill

\end{document}